\documentclass{emulateapj}
\usepackage{epsfig}
\usepackage{amsmath}
\usepackage{amssymb}
\usepackage{natbib}
\usepackage{graphicx}

\shorttitle{Gravitational Heating in Massive Galaxies}
\shortauthors{Johansson et al.}

\begin{document}

\title{Gravitational Heating Helps Make Massive Galaxies Red and Dead}

\author{Peter H. Johansson$^{1}$, Thorsten Naab$^{1}$, Jeremiah
  P. Ostriker$^{2}$}
\affil{$^1$ Universit\"ats-Sternwarte M\"unchen, Scheinerstr.\ 1, D-81679 M\"unchen, 
Germany; \texttt{pjohan@usm.lmu.de} \\
$^2$ Department of Astrophysics, Peyton Hall, Princeton, USA \\}

\begin{abstract}

We study the thermal formation history of four simulated galaxies that were
shown in \citet{2007ApJ...658..710N} to reproduce a number of observed
properties of elliptical galaxies. The temperature of the gas in the galaxies
is steadily increasing with decreasing redshift, although much of the gas has a
cooling time shorter than the Hubble time. The gas is being heated and kept
hot by gravitational heating processes through the release of potential energy from
infalling stellar clumps. The energy is dissipated in supersonic collisions of
infalling gas lumps with the ambient gas and through the dynamical capturing of satellite systems
causing gravitational wakes that transfer energy to the surrounding gas.
Furthermore dynamical friction from the infalling clumps pushes out dark matter,
lowering the central dark matter density by up to a factor of two from $z=3$
to $z=0$. In galaxies in which the late
formation history $(z\lesssim 2)$ is dominated by minor
merging and accretion the energy released $(E\sim 5\times 10^{59} \rm ergs)$
from gravitational feedback is sufficient to form red and
dead elliptical galaxies by $z\sim 1$ even in the absence of supernova and AGN feedback.

\end{abstract}


\section{Introduction}

In the standard cold dark matter (CDM) picture of galaxy formation, gas falling
into dark matter halos is shock-heated approximately to the halo virial temperature, 
$T_{\rm vir}=10^{6} \ (v_{\rm circ}/167 \rm{km s^{-1}}) \ K$ 
maintaining quasi-hydrostatic equilibrium with the dark matter component. 
\citep{1977MNRAS.179..541R,1977ApJ...215..483B,1977ApJ...211..638S,1978MNRAS.183..341W}. 
The gas will cool, losing its pressure support, and settling into a centrifugally supported disk while
conserving its specific angular momentum \citep{1980MNRAS.193..189F}. 
However, recently \citet{2003MNRAS.345..349B,2006MNRAS.368....2D} showed using
one-dimensional models that 
galaxy halos can only shock-heat infalling gas if the cooling rate
for gas behind the shock is lower than the compression rate of the infalling
gas. This criterion translates to a roughly redshift independent critical minimum mass for halos that are
able to shock-heat the infalling gas of  $M_{\rm shock}\approx 10^{11.6}
M_{\odot}$. Less massive halos are not able to support stable shocks and the
majority of their gas is accreted cold. However, in the general
three-dimensional case galaxies above the critical
shock mass can also be fed with cold gas along filaments penetrating deep
inside the hot halo \citep{2008MNRAS.390.1326O}.

\begin{figure*}
\centering 
\epsscale{1.0}
\plotone{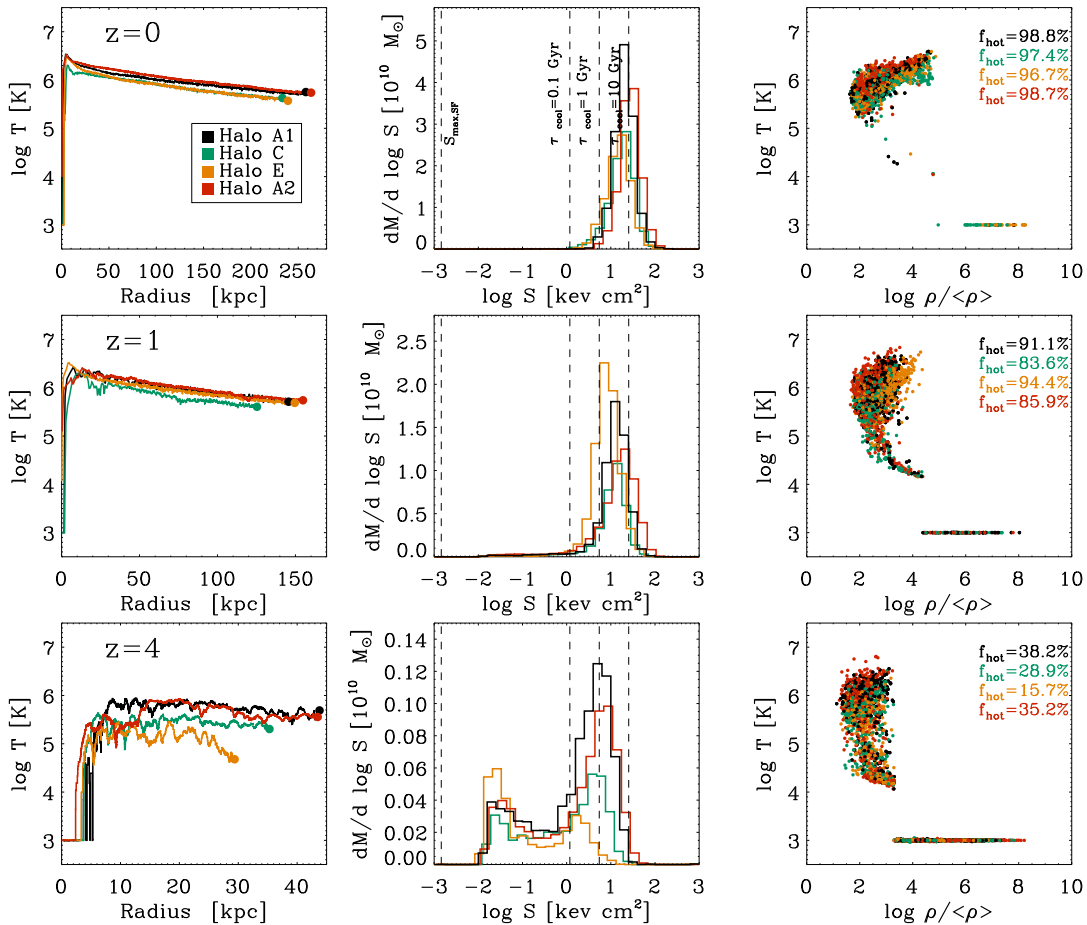}
\caption{The evolution of the temperature profile (left panel), the entropy
  distribution (middle panel) and the phase-space diagram (right panel) of all
  gas within $r_{\rm vir}$ as a
  function of redshift (top to bottom) for halos A1 (black), C (green), E
  (orange) and halo A2 (red), where $f_{\rm hot}$ is the fraction of
  gas with $T> 2.5\times 10^{5} \ \rm{K}$. 
  Even though much of the gas has a cooling time shorter than the Hubble time its temperature
  steadily increases due to the enumerated gravitational heating processes.
\label{temp_ent}}
\end{figure*}

Recent simulations also predict that the
baryonic growth of most galaxies is dominated by 
smooth accretion events of either hot or cold gas, as opposed to individual merger events 
\citep{2008ApJ...688..789G,2008arXiv0812.0007B}. 
Observations of massive star-forming galaxies at $z\sim 2$ have found a
preponderance for thick gas-rich rotating disks with large turbulent motions
and only a minority of major
mergers \citep{2006ApJ...645.1062F,2008ApJ...682..231S}. 
The observed high-redshift disks exhibit large turbulent motions that could
potentially be the result of gravitational energy release from cold accretion
flows feeding the galaxy \citep{2008ApJ...687...59G}. 
Given the bimodal temperature distribution of the accreted gas and the
characteristic mass scale of  $M_{\rm shock}$, it is tempting to link this to
the observed bimodality in the galaxy population dividing them into a red and blue
sequence with a critical stellar mass of $M_{\rm crit}\simeq 3\times 10^{10}
M_{\odot}$, which was revealed in recent large galaxy surveys (e.g. \citealp{2003ApJS..149..289B,2003MNRAS.341...33K}). 
Galaxies below this critical mass
are typically blue, star-forming disk galaxies, whereas
galaxies above $M_{\rm crit}$ are dominated by red spheroidal systems
with old stellar populations.

The observed bimodality can be explained if 
star formation in halos above a critical threshold mass of $M\sim 10^{12}
M_{\odot}$ is suppressed (e.g. \citealp{2006MNRAS.370..645B,2006MNRAS.370.1651C}). 
The quenching mechanism needs to be both energetic enough to trigger the
quenching and long-lasting enough to maintain the quenching over a Hubble
time. In addition to the quenching by shock-heated gas above a critical
halo mass \citep{2006MNRAS.368....2D,2007MNRAS.380..339B}, 
potential quenching mechanisms include the feedback from AGNs 
\citep{2007ApJ...665.1038C}, gaseous major
mergers triggering star burst and/or quasar activity 
(e.g \citealp{2006MNRAS.372..839N,2007ApJ...659..976H,2009ApJ...690..802J})
and gravitational quenching by clumpy accretion \citep{2008MNRAS.383..119D}. 
Environmental effects and the
input of gravitational energy from infalling clumps are real and automatically
included in high-resolution numerical simulations. In semi-analytic
models these environmental effects are not usually included (see, however, 
\citealp{2008ApJ...680...54K} for an exception.)

In this letter we study in detail the thermal formation history of a set of 
very high resolution cosmological re-simulations of individual
galaxies. In an earlier paper (\citealp{2007ApJ...658..710N}, hereafter N07) 
we showed that the simulated galaxies reproduce a number of observed
properties of elliptical galaxies, although the simulations include neither stellar nor AGN
feedback. We show that the gas is heated and kept hot by gravitational
feedback, efficiently terminating star formation and leading to the formation
of systems with evolved stellar populations and low star formation rates 
by $z\sim 1$. Gravitational feedback comes in many forms. Incoming
lumps or streams of cold gas ultimately come to rest depositing their
potential energy frictionally (i.e. through decaying turbulence). Supersonic
collisions of infalling gas with the ambient gas lead to propagating shock
waves \citep{2003ApJ...593..599R} which also deposit entropy throughout the system. Gradual
accretion to the gaseous envelope of the galaxy adds weight, causing
contraction and $PdV$ work on the ambient gas. And finally infalling satellite
systems captured through dynamical friction cause gaseous wakes from which
energy is transferred to the surrounding gas \citep{1999ApJ...513..252O}. These are all
variants of the processes by which gravitational energy released by infalling
matter can be transferred to the gas maintained in quasi-hydrostatic
equilibrium within galaxies, heating it and tending to counteract radiative losses.

\section{Simulations}
\label{Sims}

The $\Lambda$CDM initial conditions assume
scale-invariant adiabatic fluctuations with $\Gamma=0.2$ \citep{1992MNRAS.258P...1E}.
Throughout this letter we use a WMAP-1 \citep{2003ApJS..148..175S} cosmology with a slightly lower 
Hubble parameter of $h=0.65$\footnote{$h$ defined such that  
$H_{0}$=100$h$ kms$^{-1}$Mpc$^{-1}$.} with $\sigma_8$=0.86, $f_{b}= 
\Omega_b/\Omega_m$=0.2, $\Omega_0$=0.3, and $\Lambda_0$=0.7. The galaxies
were simulated at high resolution using the volume
renormalization technique (\citealp{1993ApJ...412..455K}) by
selecting target halos at $z=0$ in low-density environments from a 
low-resolution dark matter simulation. In the re-simulations the 
particle number of the gas and dark matter particles was increased to 
$100^{3}$ (Halos A1, C and E) and $200^{3}$ (Halo A2) within
a cubic volume at redshift $z=24$ containing all the particles that ended up
within the virialized region of the halos at $z=0$. The virial properties of
the galaxies at $z=0$ are summarized in Table \ref{gal_prop}.

The simulations were performed using the 
TreeSPH-code GADGET-2 \citep{2005MNRAS.364.1105S}
on shared-memory machines hosted in Cambridge, Munich and Princeton.
The code includes star formation and optically thin radiative cooling for a primordial
composition of hydrogen and helium \citep{1996ApJS..105...19K}.  
We included a spatially uniform redshift-dependent UV background radiation field with a modified 
\citep{1996ApJ...461...20H} spectrum.
We modify the self-regulated feedback model of \citet{2003MNRAS.339..289S} by
turning off the feedback from supernovae, eliminating the two-phase 
description of the ISM of star-forming particles with $\rho>\rho_{\rm{thresh}}$
\citep{1977ApJ...218..148M,2006MNRAS.371.1519J}, and thus transforming their entire
gas component into the cold phase with a pre-set temperature of $T_{\rm cold}=1000 \rm \ K$. 
The star formation time scale is set by
$t_{\star}=t_{0}^{\star}(\rho_{\rm{thresh}}/\rho)^{1/2}$, where we set the free
parameters to $t_{0}^{\star}=1.5 h^{-1} \rm{Gyr}$ and 
$\rho_{\rm{thresh}}= 7 \times 10^{-26} h^{2} \rm \ g \rm \ cm^{-3}$.

The gravitational softening length for the $100^3$ ($200^3$) runs  was fixed in comoving
units of $\epsilon_{\rm gas}= \epsilon_{\star}=0.25\ (0.125)
\rm \ kpc$ and $\epsilon_{\rm DM}=0.5\ (0.25) \rm \ kpc$ until $z=9$, after which the 
softening remained fixed in physical coordinates at the same values.
Our lowest (highest) resolution runs have particle masses of 
$m_{\rm stars}=1.05 \times 10^{6} M_{\odot}$ ($m_{\rm stars}=1.3 \times 10^{5}
M_{\odot}$), $m_{\rm gas}=2.1 \times 10^{6} M_{\odot}$ ($m_{\rm
  gas}=2.6 \times 10^{5} M_{\odot}$) and $m_{\rm dm}=8.2 \times 10^{6} M_{\odot}$ ($m_{\rm
  dm}=1.03 \times 10^{6} M_{\odot}$) for the stars, gas and dark matter,
respectively.

\begin{table}
\caption{Galaxy properties within $r_{\rm vir}$ at $z=0$}             
\label{gal_prop}      
\centering          
\begin{tabular}{c c c c c c c c c c}
\hline\hline       
Galaxy & $M_{\rm vir}$\footnote{Total masses $M$ in $10^{10} M_{\odot}$.} 
& $M_{\rm stars}$ & $M_{\rm gas}$ & $M_{\rm DM}$ & $r_{\rm
  vir}$\footnote{Virial radius in kpc, defined as the radius enclosing an
  overdensity of 200 times the critical density $\rho_{\rm crit}$.} & 
$v_{\rm max}$\footnote{Maximum circular velocity in $\rm{km/s}$.} \\
\hline                    
A1 & 169 & 21.7 & 13.3 &  134  & 258 & 270 \\ 
C1 & 125 & 17.2 &  9.3 &  98.5 & 233 & 258 \\
E1 & 134 & 19.0 &  9.3 &  106 &  239 & 307 \\
A2 & 179 & 22.8 & 11.6 &  145 &  263 & 232 \\
\hline                  
\end{tabular}
\end{table}

\section{Gravitational heating}
\label{grav_heating}

\subsection{Temperature structure of the gas}

We summarize the evolution of the gas temperature at redshifts $z=0,1,4$ for our model galaxies in
Fig. \ref{temp_ent} by showing the temperature profiles (left panel), the
entropy distributions (middle panel) and the phase-space diagrams (right
panel) of all gas within the virial radius.
The temperature of the gas is increasing in all galaxies with decreasing
redshift reaching $T\sim 10^{6} \ \rm{K}$ at $z=0$, with cool, star-forming gas
only found at the very centers of the galaxies. In the middle panel we plot
the distribution of the gas entropy, defined as $S=kTn^{-2/3}$ and 
define the cooling time of the gas as 
\small
\begin{equation}
t_{\rm cool}=\left(\frac{S}{10 \ \rm{keV cm^{2}}}\right)^{3/2}
\frac{1.5 (\mu_{e}/\mu)^{2}\cdot(10 \ \rm{keV cm^{2}})^{3/2}}{(kT)^{1/2}\Lambda(T,Z)},
\label{tcool}
\end{equation}
\normalsize
where $k$ is the Boltzmann constant, $\Lambda(T,Z)$ is the cooling function, 
and $\mu\simeq 0.59$, $\mu_{e}\simeq 1.1$ for a
fully ionized gas. The first term is a measure of the entropy and therefore an adiabatic
invariant, whereas the second term only depends on the temperature $(T)$ and
metalicity $(Z)$ of the gas \citep{2004ApJ...608...62S,2008MNRAS.387...13K}. 
The second factor in Eq. \ref{tcool} has an absolute minimum, corresponding to 
$t_{\rm cool}\sim 2 \ \rm Gyr$ for our primordial cooling function
(and S=10 $\rm kevcm^{2}$, with different $S$
corresponding to different $t_{\rm cool}$). We plot in
Fig. \ref{temp_ent} the entropy values corresponding to minimum cooling
times of 0.1, 1 and 10 Gyr as dashed lines. At high redshifts the entropy
distribution of the gas is bimodal with cold, high-density, star-forming gas
forming a low entropy peak and lower density, hot shock-heated gas forming a
high entropy peak. At lower redshifts the available low entropy gas has been
consumed by star formation and we are primarily left with dilute shock-heated gas with
cooling times of the order of $t_{\rm cool}\sim 5-10 \ \rm Gyr$.

\begin{figure}
\centering 
\epsscale{1.0}
\plotone{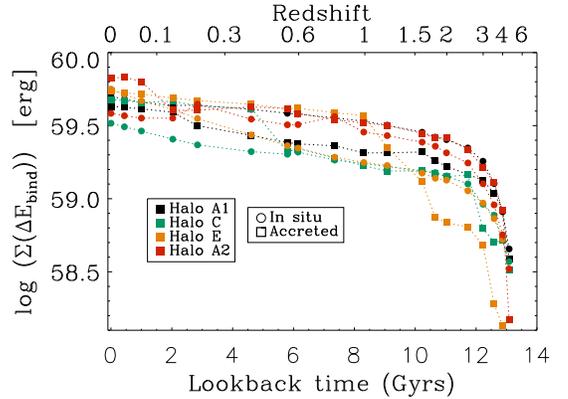}
\caption{The cumulative change in binding energy $\Delta E_{\rm bind}$
summed from z=5 to z=0 separately for in situ (circles) and accreted stars
(squares) as a function of lookback time (redshift).}
\label{grav_energy}
\end{figure}

\begin{figure*}
\centering 
\epsscale{1.0}
\plotone{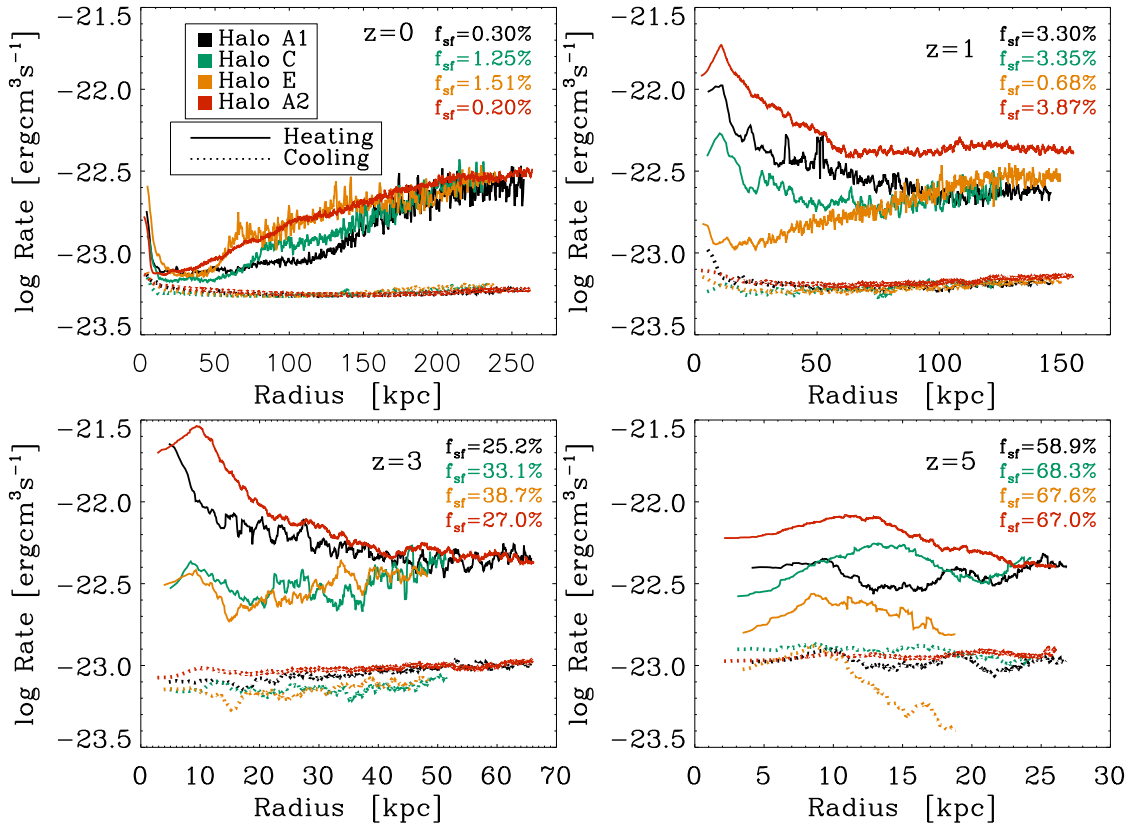}
\caption{The net heating (solid line) and net cooling rates (dashed lines) for
  non-starforming $(\rho<\rho_{\rm{thresh}})$ gas within $r_{\rm vir}$ at
  redshifts 0, 1, 3 and 5 for the four halos. The fraction $f_{\rm sf}$ of dense 
starforming gas $(\rho>\rho_{\rm{thresh}})$ is also given. The heating rate
dominates over the cooling rate at all redshifts for the low-density
non-starforming gas.}
\label{rates}
\end{figure*}

At high redshifts $(z\gtrsim 4)$ the gas flows into the galaxies
predominantly in the cold phase with the hot gas (defined as $T>2.5\times 10^{5} \ \rm
K$) fraction being low at $f_{\rm hot}\lesssim 30 \%$. A large fraction of the
gas can also be found in the cold star-forming phase. 
With decreasing redshift the fraction of hot gas steadily
increases; by $z\sim 3$ the hot gas fraction is
around $f_{\rm hot}\sim 50 \%$. The transition from a cold gas accretion mode 
to a hot gas dominated accretion mode occurs at $z\sim 2-3$, corresponding to
halo masses of $M_{\rm halo}=3-5\times 10^{11} M_{\odot}$ in good agreement
with the predictions of \citet{2006MNRAS.368....2D}, see also
  \citet{2005MNRAS.363....2K}. Below these redshifts the galaxy halos 
are massive enough to support 
stable shocks and most of the
accreted gas is shock-heated close to the virial temperature of the halos. As
this transition is mass-dependent, it occurs somewhat earlier in the more
massive halo A1, compared to the slightly lower mass halos C and E. At $z=0$ 
the hot gas fraction is $f_{\rm hot}\gtrsim 97 \%$, with only some 
residual cold star-forming gas remaining, as most of the original gas has
either formed stars or been shock-heated to high temperatures.

\subsection{Gravitational energy input}

The diffuse gas is heated by gravitational feedback in the form of 
gravitational energy release from infalling stellar and gaseous clumps. 
We quantify this effect in Fig. \ref{grav_energy} which shows the cumulative
change of the binding energies $(\Sigma (\Delta E_{\rm bind}))$ of \textit{in situ} (circles) and
\textit{accreted} (squares) stars within a fixed radius of $r<30 \
\rm kpc$. In situ stars are defined as stars that were born from gas in the
galaxy ($r<30 \ \rm kpc$), whereas accreted stars have formed outside the
galaxy and have been accreted later on. The $\Sigma_{\rm ins} (\Delta E_{\rm
  bind})$ reflects the overall mass growth and deepening of the potential
well, whereas $\Sigma_{\rm acc} (\Delta E_{\rm bind})$ describes the release
of gravitational energy from accreted structures. All of the energy released
by the growth of the potential energy of the accreted stars $(E_{\rm bind}\sim
5\times 10^{59} \rm ergs)$ goes into heating the gas and moving outwards dark
matter and previous generations of stars (on average $\sim5-10\%$ of
  the released energy ends up as kinetic energy of the stars), 
whereas much of the energy release associated with the potential 
energy growth of the in situ stars is radiated away.

All galaxies assemble rapidly at high redshifts $(z=5-3)$ with $\Sigma
\Delta (E_{\rm bind})$ of both the in situ and accreted stars increasing by an
order of magnitude. The change in the accreted $\Sigma_{\rm acc}(\Delta E_{\rm bind})$
mirrors the mass accretion history of the galaxies with the 3:1 merger for Galaxy C  
at z=0.6 and the nearly 1:1 merger for Galaxy E at z=1.5 clearly visible (see
Fig 7. in N07 for details). At $z<1$ the integrated change in the potential energy is
dominated by accreted stars for halos A1,C,A2 (dissipationless formation) and in situ stars for halo
E (dissipational formation), with the binding energy/atom increasing by $\sim20\%$ from
$z=1$ to $z=0$.The $\Sigma_{\rm acc}(\Delta E_{\rm bind})$ is
2-3 larger compared to $\Sigma_{\rm ins}(\Delta E_{\rm bind})$ for halos A1,C
and A2 at $z<1$, whereas halo E is dominated by in situ star formation and 
$\Sigma_{\rm ins}(\Delta E_{\rm bind})$ is a factor of four larger than $\Sigma_{\rm acc}(\Delta E_{\rm bind})$.
We expect gravitational feedback effects to be important in the galaxies
assembled through dissipationless processes (A1,C,A2), whereas gravitational
feedback plays a minor role in the dissipational formation of galaxy E.
We also see a trend with increasing resolution, the final $\Sigma_{\rm acc}(\Delta E_{\rm bind})$
of halo A2 is higher by 50\%, whereas the $\Sigma_{\rm ins}(\Delta E_{\rm bind})$ is
lower by 30\% compared to halo A1. Thus we expect stronger gravitational
feedback effects with increasing resolution as we are able to resolve smaller structures.

\subsection{Heating of the gas component}

The energy input from gravitational feedback scales with the mass fraction of each component, with 
roughly $\Omega_b/(\Omega_m+\Omega_b)=1/6$ and $\Omega_m/(\Omega_m+\Omega_b)=5/6$
going into the baryonic and DM components, respectively. Accreted lumps and
cold streams of gas deposit energy into the gas through dissipation of turbulence.
The infalling clumps have typical velocities of $v_{z=0}\sim 400 \ \rm kms^{-1}$, 
$v_{z=3}\sim 200 \ \rm kms^{-1}$, $v_{z=5}\sim 100 \ \rm kms^{-1}$ as estimated from the
difference of the escape velocity at $r_{\rm vir}$ and  $0.1 r_{\rm vir}$. The
corresponding sound speeds of the ambient gas are $c_{z=0}\sim 150 \ \rm kms^{-1}$, 
$c_{z=3}\sim 100 \ \rm kms^{-1}$, $c_{z=5}\sim 20 \ \rm kms^{-1}$ resulting
typically in weak shocks with Mach numbers of 2-5.

We quantify this effect by studying the terms of the entropy equation
of GADGET-2 for a given particle $i$ \citep{2002MNRAS.333..649S},
\small
\begin{equation}
\frac{dA_{i}}{dt}=-\frac{\gamma-1}{\rho_{i}^{\gamma}}
\Lambda(\rho_{i},u_{i})+\frac{1}{2}\frac{\gamma-1}{\rho_{i}^{\gamma}}
\sum_{j=1}^{N}m_{j}\Pi_{ij}\bf{v}_{ij}\cdot\rm{\nabla_{i}\bar{W}_{ij}}. 
\label{ent_evo}
\end{equation}
\normalsize
Here the first term depicts the external radiative cooling (or heating) of the gas
and the second term gives the generation of entropy by artificial viscosity in
shocks and where the internal energy is defined as $u=A/(\gamma-1)\rho^{\gamma-1}$.
In Fig. \ref{rates} we plot the cooling and heating rates
derived from Eq. \ref{ent_evo} for all non-starforming gas
$(\rho<\rho_{\rm{thresh}})$ for our four galaxies at redshifts z=0,1,3,5. The
fraction of dense starforming gas $(\rho>\rho_{\rm{thresh}})$ at $T=1000 \ \rm
K$ that is not participating in the cooling/heating process is also given in the figure.
The shock-induced heating rate of the diffuse gas is larger at all redshifts
compared to the cooling rate. At large redshifts a substantial fraction of the
gas ($\sim 2/3$ at $z=5$) and ($\sim 1/3$ at $z=3$) is starforming, whereas at
$z\lesssim 1$ this fraction is below a few percent. At $z\lesssim 1$ the
energy input from gravitational feedback is able to keep the gas hot in the
galaxies forming dissipationlessly (A1,C,A2), whereas the dissipationally
forming galaxy E exhibits much lower heating rates at $z=1$.

\subsection{Heating of the dark matter component}

\begin{figure}
\centering 
\epsscale{1.0}
\plotone{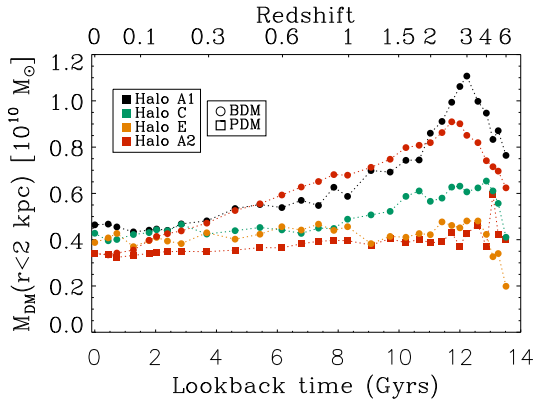}
\caption{Cumulative dark matter mass distribution within $r<2 \ \rm kpc$ 
    as a function of lookback time (redshift). The circles give the DM mass
    in our full baryons+DM (BDM) runs and the squares the DM mass in a
    pure DM (PDM) simulation of halo A2.}
\label{masscumu_dm}
\end{figure}

In addition to heating the gas the gravitational feedback heats the dark
matter (DM) of the galaxies causing it to expand outwards. We quantify this effect
in Fig. \ref{masscumu_dm}, where we plot the central DM mass within
$r<2 \rm \ kpc$ in our four galaxies (circles) as a function of lookback time.
The employed gravitational softening $(\epsilon=0.25-0.5 \rm \ kpc)$ 
is a Plummer equivalent
Spline softening length, for which the force is exactly Newtonian beyond
$h=2.8 \epsilon$. Thus $r<2 \rm \ kpc$ is resolved by $4-8\times \epsilon$,
which should be sufficient to follow the DM mass evolution reliably.

In the galaxies that form dissipationlessly we see an initial increase in the
central DM due to adiabatic contraction. After a peak DM mass at
$z=3$ the central DM mass steadily declines due to heating from infalling
clumps, with the final central DM mass being a factor of 2 lower for galaxies A1,A2
and 50\% lower for galaxy C. In contrast galaxy E shows a constant evolution
of the central DM mass, with the z=0 and z=3 central DM masses being roughly
the same. We also plot for comparison the central DM of a pure DM only simulation
of halo A2 at $200^3$ resolution (red squares). For the PDM simulation we see
neither the adiabatic contraction phase nor the pushing out of DM, thus
resulting in a constant central DM mass throughout the simulation.

This has important implications for the recent estimates of central dark
matter content of elliptical galaxies (e.g. \citealp{2009ApJ...691..770T}).
The gravitational energy release from infalling clumps might help in transforming an initially
cuspy DM profile to a cored profile as seen in our simulations 
(see also \citealp{2001ApJ...560..636E,2008ApJ...685L.105R,2009arXiv0901.1317R}). 
Furthermore, once a cored DM profile with a hot low-density gas halo is
in place, it can be maintained through the infall of cold gas clouds even
without the inclusion of additional feedback sources \citep{2008arXiv0812.2025K}.
Finally, the accretion of minor mergers can potentially explain the strong size evolution
observed in elliptical galaxies between $z=2-0$ \citep{2008ApJ...677L...5V,2009arXiv0903.1636N}.

\section{Discussion} 
\label{discussion}

In this letter we have studied the formation history of four
galaxies that reproduce a number of the observed properties for elliptical
galaxies. We have shown that for three galaxies (A1,C,A2) for which the
late formation $(z\lesssim 2)$ is dominated by dissipationless minor merging
that the energy release from gravitational feedback alone is sufficient
to make these galaxies red and dead by $z=1$, the fourth galaxy (E) has a
dissipational formation history with significant ongoing star formation at low
redshifts. Stellar and AGN feedback are both real and important, especially in driving winds and outflows that
enrich the intergalactic medium with metals
(e.g. \citealp{2006MNRAS.373.1265O}) 
and potentially in determining the lower and upper cutoffs in the observed galaxy
luminosity function \citep{2003ApJ...599...38B}. Furthermore, the baryon/DM
fraction in our simulated galaxies is a factor of two too high for their mass 
compared to predictions from recent lensing data \citep{2006MNRAS.368..715M}.
However, in order to isolate
the effects of gravitational heating from infalling clumps, which would occur
even if there were neither stellar nor AGN feedback, we chose to run our simulations
without any additional energy feedback. 
 
The efficiency of gravitational feedback can be estimated using the following
simple argument,
\small
\begin{equation}
\frac{d\phi}{dr}=-\frac{Gm_{0}}{r_{0}}\frac{1}{r}=-\frac{v_c^{2}}{r}
\Rightarrow \Delta\phi(r)=-v_c^{2}\log (\frac{r_{\rm vir}}{r}),
\label{eq_grav_heat}
\end{equation}
\normalsize
where the final change in potential energy results from integrating from 
$r_{\rm vir}$ to a fiducial infall radius $r$. Gravitational
feedback energy $(\Delta E)_{\rm grav}= \Delta m_{\star}\Delta \phi(r)$
resulting in 
\small
\begin{equation}
(\Delta E)_{\rm grav}=\Delta m_{\star}v_{c}^{2}\log(100)\sim 4.5\times10^{-6}v_{300}^{2}m_{\star}c^{2},
\label{eq_grav_heat_val}
\end{equation}
\normalsize
where $r_{\rm vir}/r=100$ and $v_{300}$ is the circular velocity in units of
300 km/s. Inserting the values for galaxy A1 at $z=0$ ($v=271 \ \rm km/s$ and
$m_{\star}=1.23\times 10^{11} M_{\odot}$) into Eq. \ref{eq_grav_heat_val}
results in $(\Delta E)_{\rm grav}\sim 8\times 10^{59} \ \rm ergs$ which is
remarkably similar to the values that can be read off Fig. \ref{grav_energy}.
More importantly, this energy is of the same order as the feedback energy from supernova II  
$(\Delta E)_{\rm SNII}\sim  2.8\times10^{-6}m_{\star}c^{2}$ (for a Salpeter
IMF and $10^{51}$ ergs per supernova) and AGN feedback energy 
$(\Delta E)_{\rm AGN}\sim  5\times10^{-6}m_{\star}c^{2}$  (for
$m_{BH}/m_{\star}=10^{-3}$ and a total AGN feedback efficiency of
0.5\%). However, unlike SNII and AGN feedback energy, gravitational feedback
energy scales with $(\Delta E)_{\rm grav}\propto v_{c}^{2}$, making it
proportionally more important in massive systems with large circular velocities.
Furthermore, the energy released from gravitational feedback is distributed
throughout the galaxy halo, whereas the feedback energy from SNII and AGN
feedback is confined to the central region of the halo in regions with high
density cold gas that can efficiently radiate away the energy.

The process presented here is general, but in detail might depend on the
implementation of additional feedback. The inclusion of supernova feedback might be important in lowering the
number of stellar clumps thus decreasing the amount of gravitational feedback
in the system. Another important factor not included in our simulations is
metal cooling that is expected to increase the cooling rates by a
factor of ten or more. However, the increased cooling rates might to some
extent be counteracted by efficient stellar and AGN feedback, especially in
the central regions of the galaxies. Finally, the derived heating rates at the
peak of the gravitational feedback phase ($z=1$ in Fig. \ref{rates}) are
typically 15-20 times higher than the cooling rate of gas with primordial
composition and thus gravitational feedback energy is expected to be an
important energy source in the 
dissipationless formation process of all massive galaxies.

\begin{acknowledgements}

%We thank the anonymous referee for a careful reading of the manuscript and
%valuable comments.
We thank G. Efstathiou for stimulating discussions.
This research was funded by the DFG cluster of excellence 'Origin and Structure
of the Universe'.

\end{acknowledgements}

\end{document}